# A dynamic classifying approach for materials

Weipeng Hu [*]

School of Civil Engineering and Architecture, Xi'an University of Technology, Xi'an,

Shaanxi, 710048, P.R. China

**Abstract**: Considering the time-varying of the poriness of the materials in the loading process, a dynamic classifying method for the materials is proposed based on the strain factor defined from the strain relationship in three orthogonal directions. Comparing with the current classifying approach based on the Poisson's ratio, the method presented in this work is a dynamic approach that can reflect the real-time strain relationship in three orthogonal directions as well as the varying specific molecular structures partly in the loading process of the materials.

## 1 Introduction

As one of the important parameters of the engineering materials, the Poisson's ratio describes the relationship between the longitudinal strain and the lateral strain of the linear-elastic materials with small strain. In a long period following the definition of the Poisson's ratio was proposed, many studies focused on the value of which reported. In 1829's, the Poisson's ratio was proved to be $1/4$ based on the interaction theory between the molecules. Unfortunately, this theoretical result isn't agreed with the results obtained in laboratory. Thus, G. Wiltham thought that the value of the Poisson's ratio is close to $1/3$ in 1848 and which was proved in laboratory widely for almost all the metallic materials. But, for vast other materials, such as layered metal-like ceramics $M_{N+1}AX_N$ and beryllium, the values of the Poisson's ratio are much lower, even the values are negative [1-5]. By now, it is accepted that the Poisson's ratio values are vary from -1 to 1/2 for different materials, regardless of natural or artificial [6,7].

---

[*] Corresponding author: wphu@nwpu.edu.cn

Theoretically, as a material constant, the Poisson's ratio should be independent of the working conditions and the specimen size. But, practically, when the value of the Poisson's ratio for a certain material is tested, the working conditions should be adjusted to the standard experimental conditions and the specimen should be prepared in the standard size, which implies that the Poisson's ratio defined currently is defective and the classification method for the materials based on the Poisson's ratio shows some limitations.

The limitations of the classification method referring to the Poisson's ratio results from the original definition of which, $v_{ij} = -\varepsilon_j/\varepsilon_i$. According to this definition, the Poisson's ratio is a time-independent dimensionless constant describing the strain relationship in two orthogonal directions, which implies that only the small strains in two orthogonal directions are considered while the strain in the third orthogonal direction is neglected. For some two dimensional materials [8-10], this definition is correct. But, for the three-dimensional materials, the strains in arbitrary three orthogonal directions are comparable and none of which can be ignored safely. Moreover, because of the neglecting of both the strain in the third orthogonal direction and the poriness change in the definition of the Poisson's ratio, the Poisson's ratio values of some materials containing the microvoid structures are negative [1-5] even in the small strain range. It is most important that the strains considered in the definition of the Poisson's ratio are assumed to be small enough to insure the time-invariance of $v_{ij}$. This assumption limits the Poisson's ratio in the small linear-elastic range and determines that the associated theories based on the Poisson's ratio are imprecise when the large/plastic strain occurs.

As a static classifying method of the materials based on the Poisson's ratio, it cannot reflect the three-direction-strain relationship of the materials dynamically during the loading process. Thus, in this work, considering the poriness change of the materials, the time-dependent strain relationship in three orthogonal directions for the material will be proposed and a new dynamic classifying approach for the materials based on which will be presented.

**2 Strain relationship in three orthogonal directions**

It is well known that the poriness of the materials (which is formulated as $P = \frac{V_0 - V}{V_0} \times 100\% = (1 - \frac{\rho_0}{\rho}) \times 100\%$, where $V_0$ and $\rho_0$ denote the apparent volume and the volume density of the specimen, $V$ and $\rho$ denote the absolutely compact volume and the density of the material respectively) will change when the strain occurs, which is shown as the variation of the apparent volume (or the volume density) of the materials directly.

In the homogeneous material, the porosity distribution is homogeneous under any stress state. With this assumption, the variation of the poriness can be completely expressed by the change of the apparent volume accompanying with any strain states for the specimen without the size and the shape limitations.

Considering the conditions of the classic uniaxial tensile test on the homogeneous material, the apparent volumes of the specimen before/after tensioned can be formulated as,

$$V_1 = l\pi r^2 \quad \text{and} \quad V_2(t) = l\pi r^2 \prod_{i=1}^{3}[1+\varepsilon_i(t)] \tag{1}$$

where, $l$ and $r$ are the length and the bottom radius of the cylinder specimen before loading respectively, $\varepsilon_i(t)$ $(i=1,2,3)$ are the strains in three orthogonal directions depending on time. (In which, $\varepsilon_1(t)$ is the strain in the loading direction. For the isotropous cylinder specimen with

small strain, $\varepsilon_2(t)=\varepsilon_3(t)$.)

It is certain that the mass of the specimen is a constant during the loading process, which implies,

$$\rho_1 V_1 = \rho_2(t) V_2(t) \tag{2}$$

that is,

$$\frac{1-p_1}{1-p_2(t)} = \prod_{i=1}^{3}[1+\varepsilon_i(t)] \tag{3}$$

where, $p_1$ and $p_2(t)$ are the poriness of the specimen before/after tensioned. Eq. (3) is the theoretical strain relationship in three orthogonal directions considering the variation of the poriness.

## 3 Classifying materials according to the above relationship

Referring to the value of the Poisson's ratio, materials have been classified into two types: one is the classic materials with the positive Poisson's ratio and another is the named negative Poisson's ratio material. It has been mentioned that only the small strains in two orthogonal directions are considered and the change of the poriness is neglected in this classification process. With the strain relationship in three orthogonal directions considering the variation of the poriness shown in Eq.(3), a new classification standard for materials can be proposed.

Defining the strain factor as $\xi = \frac{1-p_1}{1-p_2(t)} = \prod_{i=1}^{3}[1+\varepsilon_i(t)]$, the iso-surface of $\xi=1$ is shown in Fig. 1. For the strain vector $(\varepsilon_1(t), \varepsilon_2(t), \varepsilon_3(t))$ located on this iso-surface, the poriness of the associated material in this strain state is equal to that of the zero-strain state. Thus, taking this iso-surface ($\xi=1$) as the reference, the materials can be classified into the following three types (in fact, the materials may experience all of these types during the strain evolution): The first type is the materials with $\xi<1$, which is named as under-close-grained material. Under-close-grained implies that the poriness of the material will decrease in the further loading process. The second type is the materials with $\xi=1$, which is named as critical-close-grained material. Critical-close-grained implies that the poriness of the material in this strain state is equal to that of the zero-strain state. The third type is the materials with $\xi>1$, which is named as over-close-grained material. Over-close-grained implies that the poriness of the material will increase in the further loading process.

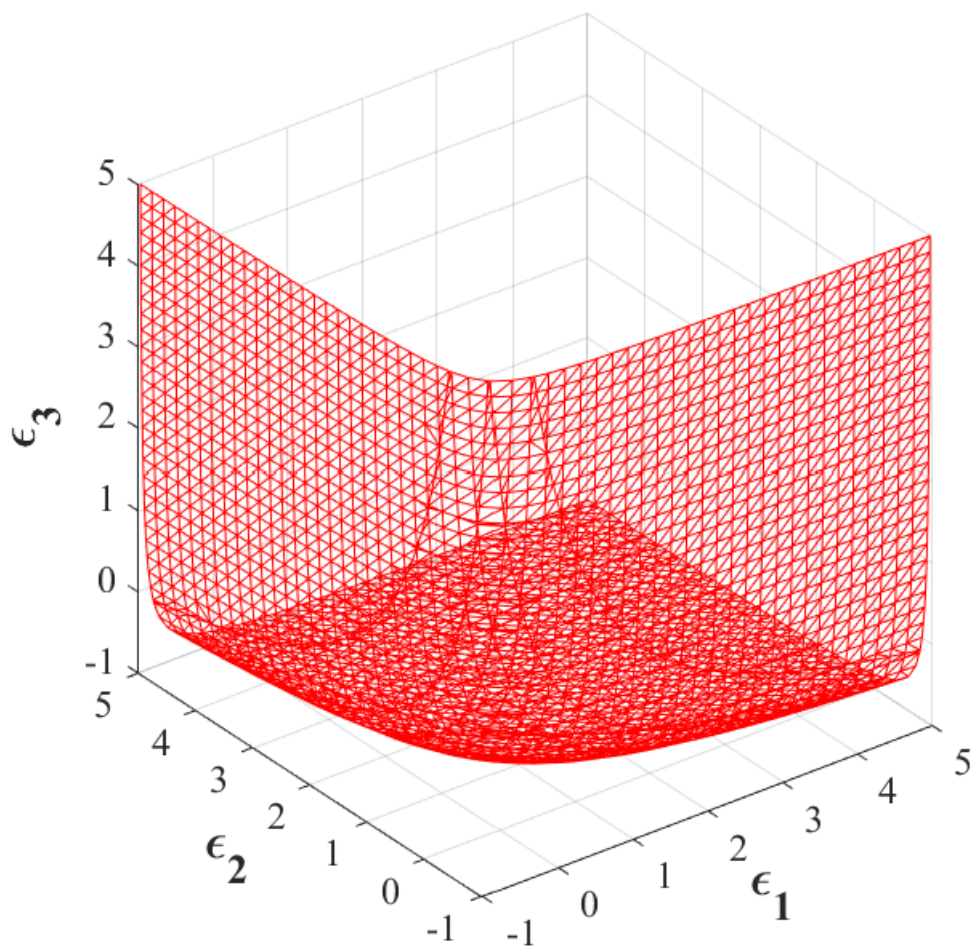

Fig. 1 The iso-surface of $\xi=1$

It needs to clarify that both the under-close-grained and the over-close-grained are independent of the absolute poriness of the materials, but depend on the poriness variation tendency during the loading process.

Classifying the materials referring to the value of the strain factor $\xi$ will end the long-term dispute on the value of the Poisson's ratio. In addition, it will give a clearer comprehension on the relationship between the strains in three orthogonal directions and the variation of the poriness.

For the classic uniaxial tensile test on the standard specimen, $\varepsilon_1(t)$ is the strain in the loading direction and the strains in other two direction are equal ($\varepsilon_2(t)=\varepsilon_3(t)$). With the small strain hypothesis, the strain factor is greater than one when the Poisson's ratio of the material is assumed as $\nu=0.3$, which implies that most of the metal materials (the value of the Poisson's ratio is close to 0.3) are over-close-grained in the small strain stage. With the increase of the strain (if the tension is in the elastic range), the strain factor of the metal material will decrease. When $\varepsilon_1 \approx 0.940363884733242$ ($\nu=0.3$), the strain factor of the metal material is close to 1, which implies that the metal material in this strain state is critical-close-grained. Subsequently, the strain factor will less than 1 if the tension process goes on and the metal material becomes under-close-grained in this situation. The above transformation process of the metal material illustrates that the classifying idea proposed in this work considers the strain property of the material dynamically.

As for the negative Poisson's ratio materials, the specimen will expend laterally accompanying with the extension in the axial direction, which means that $\varepsilon_i(t) > 0$ for $i=1,2,3$. In this situation, the strain factor $\xi=\prod_{i=1}^{3}[1+\varepsilon_i(t)] > 1$, which implies that the negative Poisson's ratio materials also belong to the over-close-grained material in the initial tensile stage. This conclusion is paradoxical to the common sense on the negative Poisson's ratio materials: Most of the negative Poisson's ratio materials are porous [1,4]. Why they belong to the over-close-grained material? It has been implied that the over-close-grained material in this work means the material owning the poriness increase tendency during the tension process. Even though the negative Poisson's ratio materials are porous, the specific molecular structures [2,4] of which permit the poriness of them to increase in the initial tensile stage. With the strain evolution, these specific molecular structures will disappear gradually and the negative Poisson's ratio materials will turn into common materials with the positive Poisson's ratio, which implies that in the following loading process, the strain factor will decrease and the materials will become under-close-grained.

## 4 Conclusions

The above analysis implies that, the classifying approach proposed in this work is dynamic, i.e., all materials will go through the transformation from the over-close-grained state to the under-close-grained during the tension process. This characteristic illustrates that the dynamic classifying method based on the strain factor is appropriate for the whole strain evolution process and can reflect the material properties dynamically.

## Acknowledgements

The research is supported by the National Natural Science Foundation of China (11672241). The author declare that he has no conflict of interest with the present manuscript.